# XML Information Retrieval:An overview

Suma D[1], U. Dinesh Acharya[2] and GeethaM[3], Raviraja Holla M[4]

[1]Department of Computer Science and Engineering, Manipal Institute of Technology, Manipal, Karnataka
India
suma.d@manipal.edu

[2]Department of Computer Science and Engineering, Manipal Institute of Technology, Manipal, Karnataka
India
dinesh.acharya@manipal.edu

[3]Department of Computer Science and Engineering, Manipal Institute of Technology, Manipal, Karnataka
India
geetha.maiya@manipal.edu

[4]Department of Information Technology, Manipal University Jaipur, Jaipur
India
raviraja.holla@jaipur.manipal.edu

*Abstract*—Locating and distilling the valuable relevant information continued to be the major challenges of Information Retrieval (IR) Systems owing to the explosive growth of online web information. These challenges can be considered the XML Information Retrieval challenges as XML has become a de facto standard over the Web. The research on XML IR starts with the classical IR strategies customized to XML IR. Later novel IR strategies specific to XML IR are evolved. Meanwhile literatures reveal development of the rapid and intelligent IR systems. Despite their success in their specified constrained domains, they have additional limitations in the complex information space. The effectiveness of IR systems is thus unsolved in satisfying the most. This article attemptsan overview of earlier efforts and the gaps in XML IR.

## I. INTRODUCTION

As the Web keeps expanding, it is increasingly difficult for the Information Retrieval (IR) systems tofind relevant information which can satisfy users' needs based on simple search queries. The users express their need using queries. In the context of World Wide Web, the query is mapped into a set of keywords (or index terms). The number of pages indexed in search engines increases as the Web expands. If the queries are short and not likely to precisely express what the user really needs, too many pages may be retrievedmay be even irrelevant to the users' needs. IR systems must also manage the representation, storage, organization of items[1] to ensure convenient access to the information relevant to the query. Further the Web IR systems use integrating approach for querying diverse data.

XML being touted the de facto standard in the Web posits IR problems over the Web to XML IR problems on the WEB. Initial research on XML IR strategies revealed customizing the conventional IR strategies re-used in the context of XML. Later novel strategies specific to the XML IR strategies evolved. The effectiveness of these strategies is environment-specific. Meanwhile the evolved intelligent and rapid IR strategies leveraged the advanced computing power.

Traditional ad hoc IR techniques are then extended yielding to personalized search and exploratory visualization techniques. Contemporarily an IR technology that integrates personalized search and exploratory visualization resulting in adaptive visualization has evolved. The integration of best of all these strategies certainly enables more sophisticated search on Web. Moreover, the "effectiveness" of IR systems remains largely unsolved. The effectiveness can be determined from the degree of transformation of IR systems from Full-or-Nil information outcome to the useful or relevant information.

## II. XML IR STRATEGIES

XML IR would be effective if the strategies therein are scalable to the increasing number of XML documents over the web. This entails appropriate strategies for query languages, representation methods, and ranking algorithms. Such efforts are continuous from both Information Retrieval and Database communities.

Approaches for accessing logically structured documents were first proposed in the 1990s [3- 6]. In the late 1990s, as XML was adopted as the standard document format, approaches for what became known as XML information retrieval were being developed (e.g. [7-9]). INEX (Initiative for the Evaluation of XML Retrieval), formed in 2002, is a yearly evaluation campaign that provides a forum for the evaluation of approaches specifically developed for XML information retrieval [10]. INEX further boosted the research in XML information retrieval. INEX provides test collections and evaluation measures, which make it possible for organizations worldwide to evaluate and compare their XML information retrieval approaches.

The goal of an XML information retrieval system is refined to focused retrieval strategies, which aim at returning document components, i.e. XML elements, instead of whole documents in response to a user query [11].

The XML IR strategies include traditional IR strategies, however customized to the context of XML and its broad use in the web. These strategies help identify the most useful XML elements to return as answers to given queries. The strategies are as follows:

i. Representation strategies for both the content and structure of XML documents.
ii. Query languages characterizing the information need with respect to both the content and structure.
iii. Ranking strategies measuring relevance and then ranking of elements for a given query.

### III. REPRESENTATION STRATEGIES

The traditional representation strategies use indexing for information representation [2]. Similarly the XML representation strategies use indexing. But XML indexing algorithms use terms-statistics at the element level. These term-statistics are termed within-element term frequency, etf, and inverse element frequency, ief. XML indexing has the additional requirement to allow retrieval of elements at any level of granularity. Also indexing must take care of "multiple" occurrence of elements intuitive with their nested structure. Otherwise using ief to discriminate between relevant and non-relevant elements will not be effective. This led to alternative means of calculating ief resulting in different representations/indices of documents. Further research is needed to best estimate ief and to decide whether the estimation strategy depends on the retrieval model and its artifacts used to rank elements, or whether the issue of nested elements actually matters. Other alternatives like aggregation and propagation also overcome the issue of nested elements with respect to the calculation of ief. It is argued [15] that even the small elements need to be indexed when a propagation mechanism is used because they might still influence the scoring of enclosing elements. Another strategy called selective indexing [16, 17] indices the element types with the highest distribution of relevant elements in past relevance data. With this strategy, a separate index is built for each selected element type (e.g., for a collection of scientific articles, these types may include article, abstract, section, sub-section, paragraph). The statistics for each index are then calculated separately. Since each index is composed of terms contained in elements of the same type (and likely comparable size), more appropriate term statistics are generated. In addition, this approach greatly reduces the term statistics issue arising from nested elements, although it may not eliminate it. After the selective indexing is done, at retrieval time, the query is then ran in parallel on each index, and the list results (one for each index) are merged to provide a single ranking across all element types and thereby list of results. The vector space model is used to rank elements in each index.

It is not yet clear which indexing strategy is the best, as obviously which approach to follow would depend on the collection, the types of elements (i.e., the DTD) and their relationships. In addition, the choice of the indexing strategy has an effect on the ranking strategy. An interesting research would be to investigate all indexing strategies within a uniform and controllable environment to determine those leading to the best performance, across, or depending, on the ranking strategies.

Indexing and ranking are two key factors for efficient and effective XML information retrieval. Inappropriate indexing may result in false negatives and false positives, and improper ranking may lead to low precisions. Shaorong Liu et al. [18] propose a configurable XML information retrieval system, in which users can configure appropriate index types for XML tags and text contents. Based on users' index configurations, the system transforms XML structures into a compact tree representation, Ctree, and indexes XML text contents. To support XML ranking, [18] proposes the concepts of "weighted term frequency" and "inverted element frequency," where the weight of a term depends on its frequency and location within an XML element as well as its popularity among similar elements in an XML dataset. The effectiveness of this system is evaluated through extensive experiments on the INEX 03 dataset and 30 content and structure (CAS) topics. The experimental results revealed that this system has significantly high precision at low recall regions and achieves the highest average precision (0.3309) as compared with 38 official INEX 03 submissions using the strict evaluation metric.

TanakornWichaiwong et al. [19] proposes a novel approach to extend the inverted index for support query processing, namely Absolute Document XPath Indexing that allows supporting and reducing the length of time on Score Sharing scheme. In terms of processing time, the system is required an average of one second per topic on INEX-IEEE and an average of ten seconds per topic on INEX-Wiki better than GPX system.

Although existing studies of XML element retrieval have already attained both effectiveness and efficiency in query processing, these studies do not consider document updates. Web documents are frequently updated, i.e., inserted, deleted, and modified. Information retrieval systems are expected to present search results based on the latest content on the Web, especially as new topics are added to documents. Without handling updates, a search system cannot find newly inserted documents and rank documents based on old information, which reduces effectiveness.

Recently, some techniques for handling document updates have been proposed. Chen et al. [20] tackle this challenge in the field of information extraction. They report the long processing time required to apply information extraction techniques to document collections when document updates occur. As a result, a delay occurs before information extraction is available on the updated documents. To shorten the delay, they propose a method to recycle intermediate results of past snapshots. Neumann et al. [21] also utilize the information of past snapshots effectively with RDF data. Ren et al. [22] preserve not only the latest graph data but also past snapshots to trace the transition of the graph. Some studies have focused on incremental updates of an inverted index [23], [24], [25], they propose data structures of indices and physical storage methods.

Atsushi Keyaki et al. [26] added a function for updating documents to the existing XML element retrieval techniques. The mainstream approach for updating an index is to construct a new index periodically from scratch while discarding the existing one. It may take a long time to retrieve updated documents if constructing a new index is costly. Incremental updates are required to shorten this delay.

Atsushi Keyaki et al. [26] gives fast and incremental updates of indices in effective and efficient XML element retrieval systems. Although Google supports fast and incremental updates with both effective and efficient query processing, Google's approach differs from [26]. Google analyzes link information in Web pages to find important pages, while [26] study utilizes text information. This technique can be applied to other structured documents apart from the Web even if they do not have link information. Term weights must be calculated during incremental updates. [26] integrates path expressions and utilizes two filters for excluding unnecessary data. The effectiveness and efficiency of this approach are evaluated through experiments with two scenarios: the static statistics case where topics rarely change, and the dynamic case in which new topics are added frequently.

IV XML QUERY LANGUAGES

The logical structure or hierarchy of XML documents contains various content and structural elements. Users may want to specify conditions to limit the search to any specific XML elements. XML query languages have been developed with the aim to express various levels of content and structural constraints. They can be classified as content-only or content-and-structure query languages.

Content-only queries make use of content constraints only, i.e. they are made of words, which is the standard form of input in information retrieval. They are suitable for XML retrieval scenarios where users do not know, or are not concerned, with the document logical structure when expressing their information needs. Although only content conditions are being specified, XML information retrieval systems must still determine what the best fragments, are i.e. the XML elements at the most appropriate level of granularity, to return to satisfy these conditions. Outside the area of XML retrieval, the surface features, i.e., anything other than content information, have been studied mainly in the context of web retrieval.

Content-and-structure query languages provide a means for users to specify content and structural information needs. It is towards the development of this type of queries that most research on XML query languages lies [12]. The structural knowledge has implicit semantics on how and why the documents are organized in a certain way. This knowledge may help the information system to retrieve the most relevant information for a user need. The usage of this structural knowledge might not only help to decide what is the best retrieval unit given a query, but it may also help to improve the effectiveness of the content oriented search. All structural information contained in the XML fragments can help the information retrieval systems to refine their content search and to decide the best retrieval unit to be returned to the user. The assumption here is that the structure exists for a reason and tells something about the document. Therefore, the structural information is discriminative and should be used for retrieval purposes. There are different aspects of the structure of documents that could give important information to a retrieval system.

Three main categories of content-and-structure XML query languages can be distinguished, namely tag-based languages, path-based languages, and clause-based languages. With tag-based queries, words in the query are annotated with a single tag name, which specifies the type of desired result elements. XSEarch is a tag-based query language [13]. Tag-based queries are intuitive, and have been used in domains outside XML information retrieval (e.g. faceted search, web search). However, they only express simple, although important and likely common, structural constraints. They cannot express, for instance, relationship (structural) constraints which may be needed for complex retrieval scenarios. The most important concept in path-based languages is the location path, which consists of a series of navigation steps characterizing movements within an XML document. Even predicates and few functions can be used with navigation steps to have more complex queries. Xpath is a path-based XML query language. NEXI, developed by INEX, consists of a small but enhanced subset of XPath. The retrieval model of XML information retrieval system implements the enhanced function and the query processing engine of XML information retrieval system implements the structural constraint. INEX is not developed to test the expressiveness of a query language for XML information retrieval, but to evaluate XML information retrieval effectiveness. Clause-based queries for XML information retrieval can be compared to SQL, the standard query language for (relational) databases. These queries are made of nested clauses to express information needs. The most prominent clause-based query languages for XML information retrieval are XQuery and XQuery Full-Text. XQuery is an XML query language that includes XPath as a sub-language, but adds the possibility to query multiple documents and combine the results into new XML fragments. The core expressions of XQuery are the FLWOR expressions. However the text search capabilities of XQuery are limited and, in addition, the result is a set of (new) XML fragments; no ranking is performed. Thus its usefulness in XML information retrieval is limited. This has led to the development of XQuery Full-Text [14]. XQuery Full-Text has been inspired by earlier query languages for searching structured text, e.g. ELIXIR, JuruXML, XIRQL. The added text search capabilities come with the introduction of the FTContainsExpr expression. XQuery Full-Text defines primitives for searching text, such as phrase, word order, word proximity, etc. FTScoreClause expressions have been introduced to allow for the specification of score variables. XQuery Full-Text does not implement a specific scoring method, but it allows an implementation to proceed as it wishes. In other words, each XQuery Full-Text implementation can use a scoring method of its choice. Therefore, an appropriate implementation of XQuery Full-Text can allow ranking of results. From a user perspective, XQuery Full-Text may be viewed as far too complex, which is one of the reasons the INEX community developed NEXI, a path-based query language with less expressiveness than a clause-based query language, as its query language. Nevertheless, XQuery Full-Text is needed in applications involving expert users, e.g. medical domain, patent industry, law.

## V. Ranking Strategies

Given an indexed collection of XML documents, the next task of an XML information retrieval system is to return for each submitted query, with or without structural constraints, a list of XML elements ranked in order of their estimated relevance to that query. In information retrieval, retrieval models are used to calculate what is called a retrieval score (usually a value between 0 and 1), which is then used as a basis to rank documents. Many of the retrieval models developed for unstructured text (document) retrieval have been adapted to XML information retrieval to provide such a score at element level.

These scores may be used to directly generate the ranked list of elements, or as input to combination strategies required for some indexing strategies in order to rank elements at all levels of granularity. For content-and-structure queries, in the context of INEX as expressed by the path-based query language NEXI, the structural constraints must be processed to provide results that not only satisfy the content, but also the structural criteria of such queries. Also, not all relevant elements should be returned as results, as they may contain overlapping content. Some processing is needed to deal with overlapping elements in result lists. Techniques that explicitly considered the document logical (tree) structure to remove overlaps usually outperformed those that did not. There is, however, the issue of speed, as the removal of overlaps is done at query time, thus requiring efficient implementations. An interesting question would be to investigate the effect of the original result list (how good it is, and how to define "goodness") on the overlap removal strategy. There are indications that a good initial result list, where good depends on the definition of relevance in the context of XML information retrieval, leads to better overlap-free result list than a less good one [27].

Established XML query languages cannot cope with the rapid growth of information in open environments such as the Web or intranets of large corporations, as they are bound to Boolean retrieval and do not provide any relevance ranking for the (typically numerous) results. The approaches such as XIRQL or XXL that are driven by techniques from information retrieval overcome the latter problem by considering the relevance of each potential hit for the query and returning the results in a ranked order, using similarity measures like the cosine measure. But they are still tied to keyword queries, which are no longer appropriate for highly heterogeneous XML data from different sources, as it is the case in the Web or in large intranets.

In such large-scale settings, both the structure of documents and the terminology used in documents may vary. As an example, consider documents about courses in computer science, where some authors talk about "lectures" while others prefer to use "course", "reading", or "class". Boolean queries searching for lectures on computer science cannot find any courses or other synonyms. Additionally, courses on database systems will not qualify for the result set, even though database systems, is a branch of computer science. So in order to find all relevant information to a query, additional knowledge about related terms is required that allows us to broaden the query, i.e., extending the query with terms that are closely related to the original query terms. However, imprudent broadening of the query may be misleading in some cases, when the extended query yields unwanted, irrelevant results. Consider a user searching for lectures on stars and galaxies. When the query to "star" is extended using related terms, the terms like "sun" and "milky way" that help in finding better results, but also terms like "movie star" or "holly wood" which are clearly misleading here. This can happen because words typically have more than one sense, and it is of great importance to choose the right sense for extending the query. Such information can be delivered by an ontology, which models terms with their meanings and relationships between terms and meanings.

Researchers in artificial intelligence first developed logic-based ontologies to facilitate knowledge sharing and reuse. Ontologies have become a popular research topic and several AI research communities such as knowledge engineering, natural language processing and knowledge representation have investigated them. In contrast to the extremely ambitious early AI approaches toward building universal ontologies, more recent proposals are aiming at domain-or user-specific ontologies and are based on more tractable logics. The publications on formalization of ontologies cover a wide spectrum from algebraic approaches and logic-based languages for modeling ontologies to ontologies for conceptual data modeling and data interpretation. Similar work has been done in the context of multi-databases, for example, proposes summary schemas model where the semantic power comes from the linguistic knowledge representation in an online taxonomy, and presents a conceptual organization of the information space. These publications do not consider the quantification of relationships.

Adding similarity measures to ontologies or, more generally, similarity measures for words has been an active research topic in linguistics for several decades. Among the first results are Rubenstein and Goodenough's judgments for semantic similarity of 65 pairs of nouns that were estimated by 51 experts. This experiment was repeated by Miller and Charles with a reduced set of 30 pairs of nouns, which has served as a basis for comparison to automatically generated similarity measures since then. The first automatic approaches for term similarity concentrated on exploiting statistical correlations between terms, especially for the problem of query expansion. Early work on similarity measures for ontologies like WordNet or other semantic nets concentrated on the graph structure of the ontology, computing similarity between two concepts by counting edges between them, considering the relative depth of the concepts in the graph or the direction of the edges between them, or taking the density of the ontology graph in to account. Since the mid of the 90ies, researchers started connecting both worlds, yielding similarity measures that take into account both the graph structure as well as statistics on a large corpus. A detailed comparison of similarity measures for WordNet can be found in [28]; [29] and [30] compare measures based on WordNet with similar measures for Roget's Thesaurus [31].

Semi-automatic or automatic ontology construction is proposed in many publications and is mostly based on methods of text mining and information extraction based on natural language processing using an existing thesaurus or a text processor such as SMES [32] or GATE [33]. Merging ontologies across shared sub-ontologies is described in [34, 35]. Some comprehensive systems for developing or using ontologies are OntoBroker, Text-To-Onto, GETESS, Prot´eg´e 2000, LGAccess, KAON, Ontolingua, and FrameNet. However the role of ontologies in searching semi-structured data has not been discussed in any depth. Although many approaches for representing and applying ontologies have already been devised, they have not found their way into search engines for querying XML data. The unique characteristic of [36] approach lies in the combination of ontological knowledge and information retrieval techniques for semantic similarity search on XML data. [36] specifies how ontologies with quantified semantic relationships can help to increase both the recall and precision for queries on semi-structured data. This is achieved by broadening the query with closely related terms, thus yielding more results, but only after disambiguating query terms, so only relevant results are included in the result of the query.

Ying Lou et al.[37] proposes a novel semantic ranking scheme SRank for XML keyword search. Keyword search is a user-friendly mechanism used to retrieve XML data for web and scientific applications. Unlike text data, XML data contain rich semantics, which are obviously useful for information retrieval. It is observed that most existing approaches for XML keyword search either do not consider relevance ranking or perform relevance ranking using traditional text IR techniques. SRank is based on a sound analysis of XML semantics and user information need. Its major contribution is the measurement of the relevance between a keyword query and an XML fragment according to their semantic similarity. The experiments demonstrated that SRank outperforms existing approaches with respect to search quality. Its efficiency scalability also bodes well for its practical implementation.

It is observed that most existing techniques for performing XML keyword searches were proposed for homogeneous XML data, such as DBLP and Wiki data. As XML and its relevant technologies (such as semantic web) have evolved, it has become clear that keyword search must be performed over heterogeneous XML data in many application scenarios. Recently, J. Pound et al. [38] proposed a method for the search problem over large-scale web-extracted data. However, further investigations are clearly necessary. Another research direction worth pursuing is to effective integration of the IR and database approaches for XML keyword search. Over the past years, the database community has initiated the LCA-based approach and the IR community has proposed the search approach based on a language model under the initiative of INEX. Determining how to integrate these two approaches for better search experience over complex and heterogeneous XML data is also an interesting subject for future work.

As the queries issued to retrieval systems are mostly short and ambiguous, they match vast numbers of documents concerning variety of subjects. Creating a linear hit list out of such a broad set of results often requires trade-offs and hiding documents that could prove useful to the user. If shown an explicit structure of topics present in a search result, users quickly narrow the focus to just a particular subset (slice) of all returned documents. Hence clustering becomes a front-end utility for not only searching but also for comprehending information. The clustering methods in such applications are inevitably connected with finding concise, comprehensible and transparent cluster labels a goal missing in the traditional definition of clustering in information retrieval. The information retrieval with XML documents is more significant to explore efficient clustering algorithms to derive comprehensible cluster labels as the XML structural and content elements are both semantic rich and hence comprehensible. Thus conflating clustering with XML might yield novel and useful information retrieval systems.

VI. SUMMARY

The information retrieval system returns a lot of documents to the information sought over the internet. Thus calculating relevance is a key factor assuring documents most likely containing answers to the query pushed up in the final ranking. XML IR systems use a very simple, but fast Boolean model of term containment to find matching documents and rely on two things to provide valuable service to the user: qualitative ranking algorithms and, most of all, the user's ability to rephrase the query until the information-need is satisfied. The challenges that make structured XML retrieval more difficult can be summarized to focused retrieval, passage retrieval, schema heterogeneity or diversity associated with its interface design as well the extended queries. Mutual compliment of clustering algorithms and XML information retrieval systems can contribute better performing XML information retrieval systems.